\documentclass[12pt,a4paper]{article}
\usepackage[scale={.75,.8}]{geometry}
\usepackage{pstricks}
\usepackage{graphicx}
\usepackage[latin2]{inputenc}
\usepackage{epic}
\usepackage{latexsym}
\usepackage{epsf}
\usepackage{epsfig}
\usepackage{cite}
\usepackage{amssymb,amsmath}
\def\ra{\rangle}
\def\la{\langle}
\def\pa{\partial}
\def\msusy{M_{\rm SUSY}}

\begin{document}

\vspace{-1truecm}

\rightline{CPHT RR-164.1107}
\rightline{LPT--Orsay 07/122}
\rightline{hep-th/yymmnnn}
\rightline{November 2007}

\vspace{0.cm}

\begin{center}

{\Large {\bf Moduli stabilization with Fayet-Iliopoulos uplift }}
\vspace{1 cm}\\

{\large E. Dudas$^{1,2}$, \ Y. Mambrini$^2$, S. Pokorski$^{3}$ and \ A. Romagnoni$^{2,1}$
}
\vspace{1cm}\\

$^1$
CPhT, Ecole Polytechnique 91128 Palaiseau Cedex, France
\vspace{0.3cm}\\

$^2$
Laboratoire de Physique Th\'eorique,
Universit\'e Paris-Sud, F-91405 Orsay, France
\vspace{0.3cm}\\

$^3$
Institute of Theoretical Physics, Warsaw University, Hoza 69, 00-681 Warsaw,
Poland
\vspace{0.3cm}\\

\end{center}

\vspace{1cm}

\abstract{
In the recent years, phenomenological models of moduli stabilization
were proposed, where the dynamics of the stabilization is essentially
supersymmetric, whereas an O'Rafearthaigh supersymmetry breaking sector is
responsible for the "uplift" of the cosmological constant to zero.
We investigate the case where the uplift is provided by a
Fayet-Iliopoulos sector.
We find that in this case the modulus contribution to
supersymmetry breaking is larger than in the previous models. A first
consequence of this class of constructions is for gauginos, which
are heavier compared to previous models.
  In some of our explicit examples, due to a non-standard
gauge-mediation type negative contribution to scalars masses,  the whole
superpartner spectrum can be efficiently compressed at low-energy. 
This provides
an original phenomenology testable at the LHC, in particular sleptons
are generically heavier than the squarks.}

\newpage

\vspace{3cm}

\newpage

\tableofcontents

\vspace{3cm}


\pagestyle{plain}

\section{Introduction}

Recently, Kachru et al. \cite{Kachru:2003aw} proposed a strategy to
stabilize the moduli in the context
of type IIB string theory orientifold, following earlier work
\cite{gkp}. The KKLT set--up
involves three steps to achieve a SUSY breaking Minkowski
vacuum, while stabilizing all moduli.
We will consider in this study a KKLT--like model where all the
complex-structure moduli are fixed by the introduction of background
fluxes for NS and RR forms. All steps except the last one (uplifting
through the addition of one anti D3-brane, analyzed in detail in
\cite{Choi:2004sx}) can be
understood within the context of an effective supergravity.
Whereas several attempts \cite{D-term} tried
to use the D-term to uplift the supersymmetric minima, it
was shown that this can work only for
a gravitino mass of the order of the GUT scale. It was however possible
to obtain TeV gravitino mass by introducing corrections to the Kahler
metric \cite{Dudas:2005vn,bhk}.
Other works insisted on the possibility of using F-terms of matter
fields in a decoupled sector to uplift the anti-de Sitter minima through
metastable vacua \cite{scrucca,F-term,dpp}.

In this note we describe a new way to obtain de Sitter space with a
TeV gravitino mass by using a Fayet-Iliopoulos (FI) model \cite{fi}
as uplift sector. The uplifting is realized through the appearance
of a non-zero $F$-term induced by the $vev$'s of matter-fields
charged under an anomalous U(1)$_X$. The F-term is directly induced
through the D-term contribution in the minimization procedure.
Moreover, the U(1)$_X$ invariance of the superpotential implies a
natural coupling between the moduli fields and the matter charged
fields under the U(1)$_X$, which changes substantially the pattern
of soft breaking mass terms compared to KKLT. The framework can be
naturally realized in orientifolds with internal magnetic fields and
is simple enough to be able to address detailed phenomenological
questions. One of the main advantages compared to previous uplifts
\cite{Choi:2004sx, F-term,dpp} is a larger contribution of the
modulus to supersymmetry breaking, which increases the tree-level
gaugino masses. Moreover, due to the details of the model mostly
related to anomaly cancelation, it is natural to introduce
messenger-like fields which realize a very particular version of the
gauge mediation proposed some time ago by Poppitz and Trivedi
\cite{Poppitz:1996xw} (see also \cite{amm}), 
in which gauge mediation contributions to
scalar masses are negative. This naturally leads us to a mixed
gravity-gauge mediation scenario, where gauge contributions of a
non-standard type \cite{Poppitz:1996xw} are generated at high scale
and compete with the gravity contributions. The resulting soft
spectrum at low-energy has new features compared to other
supersymmetry breaking schemes, in particular the spectrum is
compressed, i.e. gauginos and scalar masses have values closer to
each other than in mSUGRA, gauge mediation or the mixed
modulus-anomaly mediation. For related phenomenological analysis of
string compactifications with stabilized moduli, see e.g.
\cite{gordy}.
  The plan of our paper is the following. In Section 2 we review the
  various uplift mechanisms, insisting on the (non)decoupling of the
  sector realizing the cancelation of the cosmological constant. In
  Section 3 we define our working model, based on a FI sector with an anomalous
  $U(1)_X$ gauge symmetry, and analyze its vacuum
  structure. Section 4 presents a microscopic realization in terms of
  string orientifold models with internal magnetic fields and invoking
  stringy and spacetime instantons effects in order to obtain the main
  couplings of our model. In Section 5 we couple our supersymmetry
  breaking sector to MSSM and analyze the
  resulting superpartner spectrum at high and low-energy from the
  viewpoint of electroweak symmetry breaking. In Section 6, by using
  anomaly cancelation arguments, we enlarge our model by adding
  messenger like fields, chiral with respect to the $U(1)_X$
  symmetry. The messenger fields have a peculiar spectrum, in
  particular $Str M^2 >0$ and will generate, via gauge mediation
  diagrams, non-standard gauge contributions \cite{Poppitz:1996xw},
  which will change the low-energy spectrum in an interesting way. We
  end with some brief summary of results and conclusions. The appendix
  contain a more detailed derivation of the crucial term coupling the
  SUSY breaking sector to the modulus sector.

\section{Uplifting and decoupling}

The philosophy advocated in \cite{Kachru:2003aw} to stabilize moduli
with zero cosmological constant was to separate the process into three
steps~:

\begin{itemize}

\item Add all possible fluxes in order to stabilize most of the
(in type II, the dilaton and the complex) moduli fields. 

\item Add additional (nonperturbative in type IIB) effects in order to
stabilize the remaining (Kahler moduli in type II) moduli. The
corresponding dynamics is supersymmetric, generically generating a
negative cosmological constant.

\item Uplift the vacuum energy to zero by a source of supersymmetry
breaking which perturbes only slightly the steps above.

\end{itemize}

The last step was realized originally in \cite{Kachru:2003aw} by
adding an anti D3 brane at the end of the throat in the internal
manifold, while later on it was argued \cite{scrucca,F-term} and
explicitly shown \cite{dpp} that this can be naturally realized with
the help of a decoupled sector breaking dynamical supersymmetry in the
rigid limit. In its manifestly supersymmetric realization, by denoting
collectively $T_{\alpha}$ the moduli left unstabilized after the first
above and by $\chi_i$ the fields responsible for dynamical
supersymmetry  breaking and the uplift of the vacuum energy
\begin{equation}
K^{i {\bar j}} D_i W \overline{ D_{j} W} \ = \ 3 \ m_{3/2}^2
M_P^2 \ , \label{uplift1}
\end{equation}
the decoupling of the two sectors is symbolically described in an effective
supergravity action by writing
\begin{eqnarray}
&& W \ = \ W_1 (T_{\alpha}) \ + \ W_2 (\chi_i) \ , \nonumber \\
&& K \ = \ K_1 (T_{\alpha}, {\bar T}_{\alpha}, ) \ + \ K_2 (\chi_i,
{\bar \chi}_i ) \ . \label{uplift2}
\end{eqnarray}
The result of this decoupling is the generation of the scalar
potential of the form
\begin{equation}
V \ \simeq \ V_{\rm SUSY} (T_{\alpha}, {\bar T}_{\alpha} ) \ + \
{1 \over (T_{\alpha} + {\bar T}_{\alpha})^p} V_{\rm uplift}  (\chi_i,
{\bar \chi}_i) \ + \ {\chi_i {\bar \chi}_i \over M_P^2}  V_1
(T_{\alpha}, {\bar T}_{\alpha} ) \ + \ \cdots \ , \label{uplift3}
\end{equation}
where  the index $p$ depends on details of the uplift sector and
the term $V_1$ represents the first term in an expansion which
mixes non-trivially, due to supergravity interactions, the modulus
sector with the uplift sector. For the
case of interest $\langle \chi_i \rangle / M_P \ll 1 $, the decoupling
is very efficient and has the main consequence of perturbing very
little the supersymmetric modulus stabilization dynamics. This
reflects itself in the very small contribution of the modulus to
supersymmetry  breaking, which was estimated in \cite{dpp}, for the
case of one modulus, to be
\begin{equation}
K^{T {\bar T}} D_T W {\overline D_{\bar T} W} \sim {1 \over (T + {\bar
    T})^2}  K^{i {\bar j}} D_i W {\overline D_{\bar j} W} \simeq { 3 \ m_{3/2}^2
M_P^2 \over  (T + {\bar T})^2} \ . \label{uplift4}
\end{equation}
The small contribution of the modulus to supersymmetry breaking in
this class of uplifting mechanism has
the important outcome that generically the gauginos are much lighter
than the gravitino \cite{Choi:2004sx,dpp}. Consequently, in order to find
accurate predictions, one-loop contributions, in particular the
anomaly-mediated ones are needed, resulting in the so-called mirage
unification of gaugino masses.

It was clear from the very beginning that, while such a decoupling
renders the uplifting easy to realize, it is by no means mandatory for
the stabilization with zero vacuum energy. It is indeed conceivable to
contemplate the possibility of a sector breaking supersymmetry that,
due to various reasons, in particular gauge invariance consistency
constraints, has a non-trivial coupling to the modulus (KKLT)
sector. This non-decoupling was actually forced upon us by gauge
invariance in the D-term attempts to uplift the vacuum energy
\cite{D-term}, having as a result a very heavy gravitino mass.
Notice that in the original version with anti D3 branes
\cite{Kachru:2003aw}, a naive attempt to couple more strongly the two sectors
by increasing $V_{\rm uplift}$ results actually in  a run-away
potential which destroys the minimum.

In the next sections we provide explicit examples where this
non-decoupling is successfully realized\footnote{Recently, another
example of non-decoupled sectors was provided in the context of
heterotic strings in \cite{serone}. }. Similarly to the D-term
uplifting models, the non-decoupling is unavoidable due to gauge
invariance constraints. In the present case, due to the use of a
FI uplift sector, the novelty is the presence of a new
supersymmetry breaking source which generates a positive vacuum energy,
similar to the F-term uplifting models \cite{F-term,dpp}. As a result,
compared to (\ref{uplift4}), we get a modulus contribution to SUSY
breaking bigger than in \cite{F-term,dpp} by a factor $(4+q)/3$,
where $q$ is a $U(1)_X$ charge that will
be defined more precisely later on.



\section{The model and its vacuum structure}

Our model in its globally supersymmetric limit is a variant of the Fayet-Iliopoulos
model of supersymmetry breaking. It has two charged fields
$\Phi_{\pm}$ of $U(1)_X$ charges $\pm 1$, a constant term $W_0$ relevant,
as usual, for the supergravity generalization, and a new term
parametrized by a constant $a$ which couples $\Phi_-$ to the modulus
under consideration $T$. This last term is the main novelty and ensure
the gauge invariance of the nonperturbative superpotential term.
In the language of ${\cal N}=1$ Supergravity (SUGRA), we consider the gauge
invariant superpotential

\begin{equation}
W \ = \ W_0 \ + m \ \phi_+ \phi_- + a \ \phi_{-}^q \ e^{-bT} \ , 
\label{Eq.W}
\end{equation}

\noindent
where $W_0$ is an effective parameter coming from having integrated
out all complex structure moduli through the use of fluxes,
and a Fayet-Iliopoulos (FI) term generated in the 4D effective action
of the form\footnote{The exact form of the D-term
 depends in principle on the precise form of the Kahler metric. We take
$K= |\phi_+|^2 + |\phi_-|^2 -3 \ln(T + \overline{T})$ in what follows,
but we will comment later on about other options in the  analysis.
The term $\xi^2$ can be interpreted as $\xi^2=3/2 \delta_{\mathrm{GS}}$ if
the FI term arises from non--trivial fluxes for the gauge fields living
on the $D7$--branes.}

\begin{equation}
V_D=\frac{4 \pi}{T+\overline{T}} D^2 =
\frac{4 \pi}{T+\overline{T}}
\left(
|\phi_+|^2 - |\phi_-|^2 + \frac{\xi^2}{T + \overline{T}}
\right)^2 \ . \label{Eq.VD}
\end{equation}

\noindent
The non--perturbative potential
is either generated by Euclidean D3-branes or by gaugino
condensation \cite{Nilles:1982ik} .
The charged field $\phi_-$ restores the gauge invariance of the
nonperturbative modulus-dependent superpotential
\cite{Binetruy:1996uv,adm,Dudas:2005vv}.

Indeed, the $U(1)_X$ gauge transformations act on various fields as
\begin{eqnarray}
&&\delta V_X \ = \ \Lambda_X + {\bar \Lambda}_X \quad , \quad \delta
\Phi_i \ = \ - 2 q_i \Phi_i \Lambda_X \ , \nonumber \\
&& \delta T \ = \ \delta_{GS} \Lambda_X \ , \label{model1}
\end{eqnarray}
where $q_i$ are the charges of the fields $\Phi_i$. Gauge invariance
forces the Kahler potential for the modulus $T$ to be of the form
$K (T + {\bar T} - \delta_{GS} V_X)$. This leads in turn to the FI term
\begin{equation}
\xi_{FI} \ = \ - {\delta_{GS} \over 2} \partial_T K \ = \ {3
\delta_{GS} \over 2} {1 \over T + {\bar T}} \ \label{model2}
\end{equation}
and fixes $\xi^2 = 3 \delta_{GS}/2$.
From Eq. (\ref{Eq.W}), it
is clear that imposing $q=(2/3) b \ \xi^2$ ensures the gauge invariance of the model.
The numerical values we will be interested in what follows are
\begin{equation}
\xi \ \sim \ M_P \quad , \quad m \ll M_P \quad , \quad W_0 \ll M_P^3
\ . \label{numerical}
\end{equation}
The first requirement in (\ref{numerical}) is natural from the
string theory viewpoint, whereas the third relation is needed in
order to get $m_{3/2} \sim $TeV ; the landscape picture of string
theory could be invoked in order to achieve this
\cite{Kachru:2003aw}. The smallness of the mass term $m$ in our
model is then an outcome from a proper cancellation (uplift) of the
cosmological constant. The most natural explanation for it, in our
opinion, is in terms of stringy instanton effects recently discussed
in the literature \cite{stringy}, which can provide values $m \sim
\exp (-S_E) M_P$, where $S_E$ is the area of the euclidian brane
responsible for the mass term.

From Eq. (\ref{Eq.W}) we can deduce explicitly the F-part of the scalar
 potential given by

\begin{equation}
V_F=e^{K}
\left(
K^{i\overline{j}} D_iW \overline{D_j W} \ - \ 3 \ |W|^2
\right) \ ,
\label{Eq.VF}
\end{equation}

\noindent where $K^{i\overline{j}}$ is the inverse of the $K_{i\overline{j}}
=\partial^2K/\partial \Phi_i \partial \overline{\Phi_{\bar j}}$ metric and
$D_i$ is the Kahler covariant derivative : $D_i W = \partial_i W + (\partial_i
K) W$. Using a conventional Kahler potential of the form
$K= |\phi_+|^2 + |\phi_-|^2 -3 \ln(T + \overline{T})$, we can rewrite
Eq. (\ref{Eq.VF}) as\footnote{The Kahler metric of the charged fields
  $\Phi_{\pm}$ can be more complicated and can also depend on $T$. We
  checked that the results do not change significantly when a more
  general Kahler potential is considered.} :

\begin{equation}
V_F=\frac{1}{(T+\overline{T})^3}
\left[
\frac{(T+\overline{T})^2}{3}|W_T - \frac{3}{T + \overline{T}} W|^2
+|D_+ W|^2 + |D_- W|^2 -3 |W|^2
\right] \ . \label{Eq.VFbis}
\end{equation}

The scalar potential is given explicitly as 

\begin{eqnarray}
V (\phi_+,\phi_-, T)&=& \frac{1}{\left( 2 Re[T]\right)^3} \Bigg[
\frac{\left( 2 Re[T]\right)^2}{3} |a b \phi_-^{q} e^{-bT} |^2
\nonumber \\
&& + 2 Re[T] \left(  a b \phi_-^{q}  e^{-bT} \bar{W} + \bar{a}
\bar{b} \bar{\phi_-}^q e^{-\bar{b} \bar{T}} W\right)
\nonumber \\
&& + |m \phi_+ + a q \phi_-^{q-1} e^{-bT}  + \bar{\phi}_- W |^2 + |m
\phi_- + \bar{\phi}_+ W |^2
\Bigg] \nonumber \\
&+& \frac{4 \pi}{2 Re[T]} \left[ |\phi_+|^2-|\phi_-|^2 +
\frac{\xi^2}{2 Re[T]} \right]^2 . \label{Veff}
\end{eqnarray}
\noindent
The nonperturbative term has significant consequences both on the
resolution of the equation of motion for $\phi_+$ and $\phi_-$, and
on the uplifting mechanism. It is important to notice here that due
to the intricate coupling between $\phi_-$ and $T$, in solving the
equations of motions, we cannot neglect the supergravity corrections
to $|F_+|^2$ and $|F_-|^2$ in the scalar potential. In what follows
we define as usual
\begin{equation}
F^i \ = \ e ^{K/2} \ K^{i {\bar j}} \ \overline{ D_j W}  \ .
\end{equation}

Asking for a zero cosmological constant at the minimum, we can find
immediately a relation at first order between the gravitino mass and
the parameters of the model by anticipating that the uplift is
mainly induced by $F_+$ :
\begin{equation}
|F_{+}|^2 \ \simeq \ 3 m_{3/2}^2 M_P^2 \quad \rightarrow \quad   |m
\phi_-|
\ = \ \sqrt{3} \ |W_0| \ . \label{upliftW0}
\end{equation}

\noindent

Solving now the equations $\partial V_{eff} / \partial \phi_+=
\partial V_{eff} / \partial \phi_-=0$, using the approximations
allowed by the choice of the parameters, and always fixing the
cosmological constant to zero, we obtain at the first order
\footnote{For simplicity, we take all the parameters to be real and
we choose the real positive solution for the vev of $\phi_-$. The
general case of complex parameters does not change significantly the results.}
\begin{eqnarray}
D &=&|\phi_+|^2 - |\phi_-|^2+\frac{\xi^2}{2 Re[T]} = \frac{2
Re[T]}{8 \pi} \frac{m^2}{(2 Re[T])^3} \ .
\label{D} \\
\phi_- & = & \sqrt{\frac{\xi^2}{2 Re[T]}} \ = \ \sqrt{\frac{3 q}{4 b
Re[T]}}
 \ , \label{phimoins} \\
\phi_+ & = & -\frac{3 q}{4 b Re[T]} \left[ \frac{a q e^{-b T}}{2
m}\left( \frac{3 q}{4 b Re[T]} \right)^{\frac{q-3}{2}}
-\frac{1}{\sqrt{3}} \right] \ . \label{phiplus}
\end{eqnarray}

We can check our approximation by defining a parameter $\tilde
\epsilon$ which will be fundamental in the calculation of the soft
breaking term. $\tilde \epsilon$ measures the contribution of $T$ to
the uplift :

\begin{equation}
\tilde \epsilon \ = \ 2 Re[T] \ \frac{a b e^{-b T} \phi_-^q+3 W/(2
Re[T])}{\sqrt{3} m \phi_-} \ = \ \frac{F_T}{F_{+}} \ .
\end{equation}

\noindent
Solving $\partial_T V(T,\phi_+,\phi_-)_{\phi_+,\phi_-}=0$
with the reasonable hypothesis\footnote{ If the conditions
(\ref{hypothesis}) are violated, it turns out not to be possible to
realize the uplift of the cosmological constant with a TeV gravitino mass.}
\begin{equation}
a \ e^{-bT} \ \ll \ W_0 \ \ll m \label{hypothesis}
\end{equation}
and $\phi_+ \ll \phi_-$ (hypothesis that we check aposteriori),
we obtain at first order

\begin{equation}
\tilde \epsilon \ = \ \frac{4+q}{2 b Re[T]} \ - \
\frac{2}{\sqrt{3}}\phi_+ \ , \label{effective1}
\end{equation}

\noindent which gives for a typical KKLT value $2 b Re[T]=60$,
$\tilde \epsilon(q=1)\sim \frac{1}{12}$ and $\tilde \epsilon(q=2)
\sim \frac{1}{6}$ which is bigger than the values obtained in
sequestered F-term uplifting \cite{dpp}, where
\begin{equation}
{\tilde \epsilon}_{{\rm F-uplift}} \ = \ {3 \over {2 b Re[T]}} \ .
\label{effective2}
\end{equation}

It turns out that the numerical solution  of the equation of motion
for $T$ is very close to the supersymmetric minimum for $T$, whereas
the numerical solutions for $\phi_+$, $\phi_-$ are very close to the
analytical ones (\ref{phimoins}), (\ref{phiplus}) ; the deviation
from it can be parameterized by expanding in a perturbative
parameter of the theory (which is $a e^{-b T}/m$ in our specific
case), checking in the meantime the analytical consistency of the
whole procedure. In fact, this procedure is remnant of the one used
in the original KKLT paper where the authors noticed that the term
induced by the anti D3--brane which is proportional to $1/Re[T]^2$,
does uplift the potential without disturbing significantly the shape
and the value of $Re[T]$ at the minimum. F--term uplifting exhibit
similar features in the sense that it can be seen (see Eq.
(\ref{Eq.VFbis})) as an uplift proportional to $|D_i W|^2/Re[T]$.
However, it is important to point two main differences with KKLT
models \cite{Kachru:2003aw} and F-term uplifting ones \cite{F-term,dpp}.
Indeed, firstly the F-term breaking parameters $F_i$ are induced by
the D-term, which imposes a non vanishing vev for $\phi_+$ and
especially $\phi_-$ at the minimum of the potential. Secondly, the
gauge invariant term $a e^{-b T} \phi_-^q$ in $W$ imposes more
constraints on the parameter space, linking directly the $F_T$ and
$F_+$ in the minimization procedure. It turns out that $F_T$ is more
important in this case and participate more to the cosmological
constant cancelation. One of the main consequences appears on the
gaugino masses ($M_i \propto F_T/(2 Re[T])$), which are heavier than
in previous uplift schemes.
 One of the main difference
with the models inspired by D-term uplifting is the possibility to
achieve a TeV scale SUSY breaking.

At the first order, the value of $T$ at the minimum respects the
condition $F_T=0$, i.e.

\begin{equation}
a b e^{-bT} \phi_-^q \ = \ - \frac{3 W}{2 Re[T]} \ \simeq  \ - \
\frac{3 W_0}{2 Re[T]} \ . \label{FT0}
\end{equation}

\noindent The mass of the gravitino is given by $W/(2 Re[T])^{3/2}$.
To illustrate the procedure, we apply the minimization condition to
find a phenomenological viable point in the parameter space. We fix
$W_0$, $b$ and $q$. $\xi^2$ is given by the gauge invariance
constraint, $t=Re[T]$ (and $m_{3/2}$) are obtained by the
minimization procedure, whereas $m$ is fine-tuned to ensure a zero
cosmological constant. For the numerical values provided in Fig. 1,
we obtain :

\begin{eqnarray}
&& m_{3/2}= 3.3~ \mathrm{TeV}, ~~~~~~ \sqrt{D} = 22.5 ~\mathrm{TeV},
~~~~~~ t= 59.4 ~ M_P \ , \nonumber \\
&& \phi_+ = -1.4~10^{-2} ~ M_P , ~~~~~~ \phi_- = 0.16 ~
M_P \ . \label{num}
\end{eqnarray}

\noindent Concerning the contribution of various fields to the
uplift, we obtain $F_T \sim F_-$ and ${\tilde \epsilon} = F_T/F_+ \sim  1/12 $.

\noindent

\begin{figure}
    \begin{center}
\centerline{
       \epsfig{file=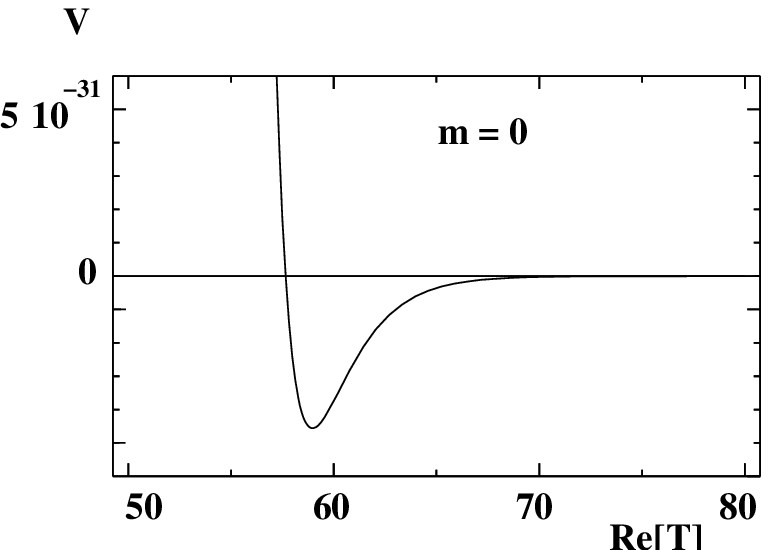,width=0.5\textwidth}
\hskip 1cm
       \epsfig{file=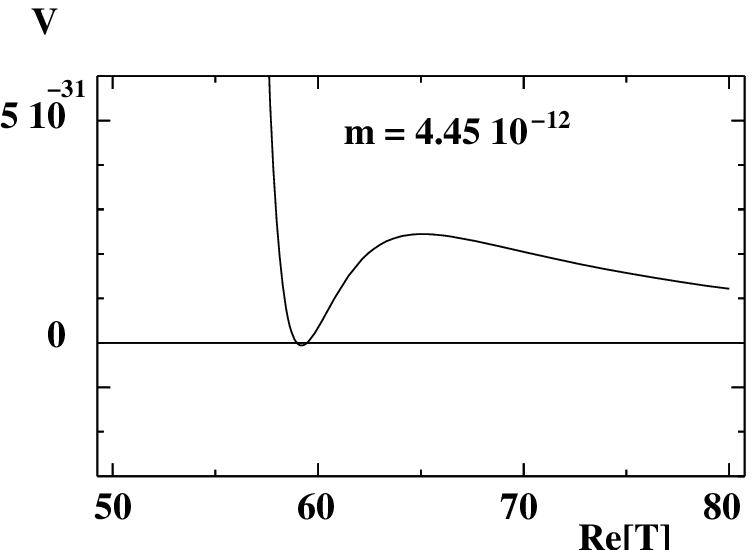,width=0.5\textwidth}
       }
          \caption{{\footnotesize
Scalar potential for $m=0$ (left) and $m \not=0$ (right) for
$W_0=-4.3 \times 10^{-13}$, $q=1$ and $b=0.5$. The other parameters are
determined by gauge invariance conditions, minimization of the
potential and zero cosmological constant : $m = 4.45 \times 10^{-12}$,
$t=59.4$ and $\xi^2=3$, which gives a gravitino mass of 3.3 TeV. }}
\label{fig:Vm}
    \end{center}
\end{figure}
\section{Microscopic definition of the model}

The setup we are considering is very similar to the one proposed in
\cite{Burgess:2003ic, Dudas:2005vv,D-term}, with slight modifications.
We start with type IIB string propagating on a Calabi-Yau manifold,
orientifolded with an involution $\Omega' = \Omega \sigma$, $\sigma^2
=1$ which generate non-dynamical O7 and O3 orientifold planes. They ask for
consistency the introduction of D7 and D3 branes. The non-trivial
dynamics we will be concerned happen on the D7 branes. The relevant
ingredients for our discussion are two stacks of D7$^{(1)}$
and D7$^{(N)}$  branes , giving rise to an $U(1)_X$ and an $U(N)$ gauge
groups. The stack D7$^{(N)}$ wraps a four-cycle of volume $V$, which suitably
combined with an axion obtained by wrapping the RR four-form over the
four-cycle $a \sim \int C^{(4)}$, forms the complex Kahler modulus $T
= V + i a$. The massless chiral open string spectrum for an arbitrary
number of stacks of branes can be given a more geometrical
interpretation by performing three T-dualites in a  IIA setting with
intersecting D6$^{(a)}$ branes \cite{intersecting}.
In IIA orientifolds with D6 branes at angles, each stack $D^{(a)}$,
containing $M_a$ coincident branes, has
a mirror $D^{(a')}$ with respect to the O6 planes. The chiral spectrum
for type II orientifold Calabi-Yau compactification with intersecting
branes contains chiral fermions in
\begin{eqnarray}
&& {\rm sector} \quad \quad \quad {\rm representation} \quad \quad \quad {\rm multiplicity \ of \ states} \nonumber \\
&& D^{(a)}-D^{(b)} \quad \quad \quad ({\bar M_a}, M_b) \quad \quad \quad \quad
\quad \quad I_{ab} \nonumber \\
&& D^{(a')}-D^{(b)} \quad \quad \quad (M_a , M_b) \quad \quad \quad \quad \quad \quad I_{a'b} \nonumber \\
&& D^{(a')}-D^{(a)} \quad \quad \quad {M_a (M_a-1) \over 2}  \quad
\quad \quad \quad {1 \over 2}
( I_{a'a} + I_{Oa}) \nonumber \\
&& D^{(a')}-D^{(a)} \quad \quad \quad {M_a (M_a+1) \over 2}  \quad
\quad \quad \quad {1 \over 2} ( I_{a'a} - I_{Oa}) \ , \label{m1}
\end{eqnarray}
where $I_{ab}$ is the intersection number between the stacks $D^{(a)}$
and $D^{(b)}$,  $I_{a'b}$ is the intersection number between the images $D^{(a')}$
and $D^{(b)}$, whereas $I_{Oa}$ is the intersection number between the
stack $D^{(a)}$ and the O6 planes. In the original type II language,
the intersection numbers are mapped into magnetic fluxes
\cite{magnetic,intersecting}.
\\
  For the two stack case discussed above and in the type IIA picture,
  we take the $U(1)_X$ brane to intersect along a six-dimensional subspace with
  the O-planes. This means that the spectrum of the states stretched
  between the $U(1)_X$ brane and its image is non-chiral and is
  described in four-dimensional language by fields $\phi_{\pm}$ of
  $U(1)_X$ charges $\pm 2$. We take the multiplicity of these states,
  which correspond to the symmetric representations in (\ref{m1}), to
  be equal to one. If the second stack $U(N)$ does not intersect the
  O-planes, the symmetric/antisymmetric representations are absent and
  only the byfundamental chiral multiplets $Q = (N,1)$ and ${\tilde Q}
  = ({\bar N},1)$, of multiplicity $N_f < N$, are charged under the non-abelian gauge group.
In the type IIB language, the axion field coupling to the $U(N)$ gauge
fields get charged under the $U(1)_X$ gauge field of the first stack if
the two stacks intersect over a two-dimensional cycle on which the
magnetic flux is non-trivial \cite{vanproeyen}. In this case we get
the typical Stueckelberg couplings
\begin{equation}
{1 \over 2} (\partial_{\mu} a + \delta_{GS} A_{\mu})^2 \ , \label{m2}
\end{equation}
rendering the $U(1)_X$ gauge field massive. The supersymmetric
description of this phenomenon is precisely the one described in eqs. 
(\ref{Eq.VD})-(\ref{model2}).

 
 If $N_f < N$, the non-abelian stack $U(N)$ will undergo gaugino
condensation and generate a non-perturbative ADS type superpotential
in terms of the "mesonic" fields $M = Q {\tilde Q}$.
\begin{equation}
W_{np} \ = \ (N-N_f) \ \left({e ^{- 2 \pi T} \over det M}\right)^{1 \over N-N_f} \ . \label{m5}
\end{equation}
It was shown long ago in a similar heterotic context
\cite{Binetruy:1996uv} and updated recently for orientifolds
\cite{Dudas:2005vv,D-term,vanproeyen} that, once the Green-Schwarz
anomaly cancelation conditions are imposed, the $U(1)_X$ charges of the
mesons are precisely such that the gauge variation of $T$ in
(\ref{m5}) is compensated by that of the mesons. In addition to the
D-term potential (\ref{Eq.VD}) and the nonperturbative term (\ref{m5}),
the other terms in the superpotential defining our model are
\begin{equation}
W_1 \ = W _0 \ + \ \lambda \ \phi_{-} Q {\tilde Q} \ + \ m \phi_{+}
\phi_{-} \ . \label{m6}
\end{equation}
The constant $W_0$ can be generated by closed string three-form fluxes
\cite{gkp,Kachru:2003aw} which stabilize the dilaton and the complex
structure moduli, whereas the second term in (\ref{m6}) is a
disk-level perturbative open string coupling. The last term, which
will turn out to be crucial for our purposes, deserves a special
discussion.  Unless the $U(1)_X$ stack and its image are parallel to
each other in some internal subspace, the mass $m$ cannot have a
perturbative origin (like for example Wilson lines). In what follows
we will advocate a non-perturbative origin $m \ll M_P$. There are two
possibilities for generating an exponentially small mass term $m$.
The first option is provided by
stringy instanton effects \cite{stringy}. The instantons under
consideration can be E$(-1)$ instantons or E3 instantons wrapping cycles
different than the one defining the Kahler modulus $T$ under
consideration. The resulting parameter m is then proportional to $m
\sim \exp (-S_E) M_P$, where $S_E$ is the instanton action.  
The other option uses a second sector undergoing spacetime nonperturbative
dynamics. This could
arises, for example, if the $U(1)_X$ brane is part of a bigger stack of
branes $U(M) = U(1)_X \times SU(M)$, with the non-abelian part $SU(M)$
undergoing non-perturbative phenomena, for example gaugino
condensation $\langle \lambda \lambda \rangle = \Lambda_M^3$. Then an
open string perturbative coupling
\begin{equation}
\int d^2 \theta \ {W^{\alpha,M} W_{\alpha,M} \over M_P^2} \ \phi_{+}
\phi_{-} \ \rightarrow \ {\Lambda_M^3 \over M_P^2} \ \int d^2 \theta \ \phi_{+}
\phi_{-} \ , \label{m7}
\end{equation}
generates a hierarchically small mass parameter $m =(\Lambda_M^3 /
M_P^2) $.
Whereas it is fair to say that constructing a complete, global model
along these lines could be a difficult task, there is no conceptual
obstruction to the implementation of the ingredients that we need in order to define
completely our model in a semi-realistic compactification.

 Finally,  by invoking the stringy
 instanton effects described previously or, alternatively, by
 integrating out the quarks $Q,{\tilde Q}$ of the hidden
 sector as described in detail in the Appendix,  we arrive at the generic
 form of the superpotential

\begin{equation}
W \ = \ W_0 \ + \ m \phi_+ \phi_- + a \ \phi_{-}^q \ e^{-bT}  \ , \label{m8}
\end{equation}
that was defining our model analyzed in the previous Section. 

\section{Soft--breaking terms}

In what follows we investigate the effects of supersymmetry breaking
in the observable sector, that we take for simplicity to be the
Minimal Supersymmetric Standard Model (MSSM). Irrespective on which
type of brane MSSM sit (D7 or D3 branes), if they contain magnetic
fluxes the gauge kinetic functions contain a T-dependence
\begin{equation}
f_a \ = \ {c_a \over 4 \pi} T \ + \ f_a^{(0)} \ , \label{sbt1}
\end{equation}
where $c_a$ are positive numbers\footnote{In the rest of the paper we
consider $c_a > 0$. This is easier to obtain in a string setup
and also safer for phenomenological purposes, since for $c_a < 0$
there is a serious danger of destabilizing the vacuum.} and $f_a^{(0)}$ effective constants
generated by the couplings of the MSSM branes to other, stabilized
fluxes (e.g. the dilaton $S$). By denoting in what follows by $i,j$
matter fields and by greek indices $\alpha$ any field contributing to
SUSY breaking, a relevant quantity for computing the soft terms
\cite{gg} is
the coupling of the matter fields metric $K_{i {\bar j}}$ to the SUSY
breaking fields. This can in turn be parameterized as
\begin{eqnarray}
&& K_{i{\bar j}}=(T+\overline{T})^{n_i}
\left[
\delta_{i{\bar j}} + (T + \overline{T})^{m_{ij}}|\phi_+|^2 Z'_{i\overline{j}}
+ (T + \overline{T})^{p_{ij}}|\phi_-|^2 Z''_{i \overline{j}}
\right. \nonumber \\
&& \left. + (T + \overline{T})^{l_{ij}}(\phi_+ \phi_-
  Z'''_{i\overline{j}} + \mathrm{h.c})
+ O(|\phi_i|^4)
\right], \label{sbt2}
\end{eqnarray}

\noindent
where $G=K+\log|W|^2$, $K_{i{\bar j}}=\partial_i \partial_{\bar j} K$, $i$ and $j$
representing the matter fields, not participating to the SUSY
breaking mechanism ($G_i=0$).  The metric $K_{i {\bar j}}$ in (\ref{sbt2})
is written as an expansion in powers of the charged vev fields $\phi_{\pm}/M_P
\ll 1$, up the quadratic order.

\subsection{Scalar masses}

For the calculation of the scalar mass, we use the classical formulas
at the linear order in the D-term \cite{Dudas:2005vv,bim}
\begin{equation}
{\tilde m}_0^2|_{i{\bar j}} \ = \ m_{3/2}^2
\left[
G_{i{\bar j}} - G^{\alpha} G^{\overline{\beta}}
 R_{i \overline{j}\alpha \overline{\beta}} \right] \ + \ \sum_a g_a^2 D_a
 \partial_i \partial_{\bar j} D_a  \ ,
\end{equation}

\noindent
with the standard definitions

\begin{equation}
R_{i \overline{j} \alpha \overline{\beta}} \ = \
\partial_i \partial_{\overline{j}} G_{\alpha \overline{\beta}}
-\Gamma_{i \alpha}^m G_{m \overline{n}}
\Gamma^{\overline{n}}_{\overline{j}\overline{\beta}}
\quad , \quad
\Gamma_{i \alpha}^m \ = \ G^{m \overline{k}}\partial_{\alpha}G_{i
  \overline{k}} \ .
\end{equation}

For the uncharged\footnote{We anticipate, for reasons that we discuss
later on, that the MSSM fields are neutral with respect to the
$U(1)_X$ symmetry.} scalar mass terms we obtain, after normalization of the
kinetic terms :

\begin{eqnarray}
&& ({\tilde m}_0^2)_{i \overline{j}}=m^2_{3/2} \left[ \delta_{i
{\bar j}} + \frac{n_i}{(T + \overline{T})^2} |G^T|^2\delta_{i{\bar
j}} -|G^+|^2(T + \overline{T})^{m_{ij}+{n_i-n_j \over 2}}
Z'_{i\overline{j}} \right. \nonumber \\
&& \left.
-|G^-|^2(T + \overline{T})^{p_{ij}+{n_i-n_j \over 2}} Z''_{i\overline{j}}
\right] \ . \label{smasses}
\end{eqnarray}
Notice that the contribution to the scalar masses coming from the
moduli, depending on the unknown moduli weights $n_i$, is suppressed
compared to the universal first term. This  comes actually from the
uplift field $\Phi_+$, via the purely supergravity interactions, as
in the mSUGRA case . The third term, coming from the main uplift
field $\Phi_+$, is also negligibly small if $r_{ij} \equiv
{m_{ij}+{(n_i-n_j) / 2}} \leq -1$, whereas it is comparable to the
universal contribution for $r_{ij}=0$ and dominant for $r_{ij} > 0$.
Whereas this last case cannot arise in a string compactification,
the case  $r_{ij}=0$ could and deserve a more detailed study from
the viewpoint of possible flavor-dependent $\Phi_+$ couplings. Since
$\Phi_+$ is a charged and therefore open-string/brane-localized
field, whereas the modulus $T$ is a closed/bulk field, the pattern
of the flavor dependence of their respective couplings to MSSM
fields is clearly different. In particular, whereas it is very
difficult to supresss the mixed modulus-MSSM fields couplings (first
term in the rhs of (\ref{sbt2})) in the Kahler potential, this can
be easily realized for the uplift open field (the second term in the
rhs of (\ref{sbt2})) $\Phi_+$\footnote{These comments also apply to
a generic F-term uplift \cite{F-term,dpp}.}. In this last case (or
if the $\Phi_+$ couplings are flavor-universal), the scalar masses
(\ref{smasses}) do not generate dangerous FCNC effects. In
conclusion, under reasonable assumptions, the dependence of the soft
masses on the unknown quantities $n_i, m_{ij}, Z'_{i{\bar j}},
Z^{''}_{i{\bar j}}$ is weak and can be neglected in a first
approximation. For the phenomenological analysis performed in the
next section we analyze in detail the universal case, where the
gravity-mediated contributions are dominated by the universal term
$({\tilde m}_0^2)_{i {\bar j}} \simeq m_{3/2}^2 \delta_{i {\bar
j}}$.

\subsection{Gaugino masses}

The gaugino mass for a general gauge kinetic function $f_a$ is given by \cite{bim}:

\begin{equation}
M_a=
\frac{\partial_T f_a}{Re[f_a]} e^{K/2} K^{T \overline{T}} D_T W \ .
\end{equation}

\noindent
With the hypothesis of a gauge kinetic function given in
(\ref{sbt1}), we obtain

\begin{equation}
M_a = m_{3/2} \alpha_a \frac{(T+\overline{T})}{3} \ \frac{D_T W}{W}
 = m_{3/2} \alpha_a \frac{(T+\overline{T})}{3} \ G_T \ , \label{gmasses}
\end{equation}
where
\begin{equation}
\alpha_a \ = \ {c_a \over c_a + 4 \pi f_a^{(0)} / T} \ .
\end{equation}
 For the phenomenological analysis performed in
the next section we analyze in detail the unified case $\alpha_a
\simeq 1$, which is easily realized for $4 \pi  f_a^{(0)} \ll c_a
T$.
\subsection{Trilinear couplings}

The general formulas for the trilinear couplings
including the D-term contribution can be found in \cite{Dudas:2005vv,bim}
\begin{equation}
A_{KLM}= e^G
\left[
3 + G^{\alpha} \nabla_{\alpha}
\right]\nabla_K \nabla_L \nabla_M G
+ \sum_a g_a^2 D_a \nabla_K \nabla_L \nabla_M D_a \ ,
\label{Ageneral}
\end{equation}

\noindent where $\nabla_i G = \partial_i G = G_i$, $\nabla_i G_j=
G_{ij} - \Gamma^k_{ij} G_k $, etc. It is easy to show that the last
contribution in (\ref{Ageneral}) coming from the D-term is in our
case negligible. Applying it to our special case, after
normalization of the kinetic terms and for a typical superpotential
for matter fields of the form $W_{m}=\frac{1}{6} W_{KLM} Q_K Q_L
Q_M$, we get

\begin{eqnarray}
A_{KLM}&=& m_{3/2} \left[
3 W^0_{KLM} - \frac{G^T}{2(T + \overline{T}) }(n_K + n_L +n_M-3) W^0_{KLM}
\nonumber
\right.
\\
&+& G^T \partial_T W^0_{KLM} - 3 G^+ {\bar \phi}_+
\left.
\left(
 (T+\overline{T})^{{n_K-n_i \over 2} +m_{Ki}}Z'_{Ki}W^0_{iLM}
\right)_{\rm symm.} \right. \label{A-terms}
\\
&-&
\left. 3 G^-  {\bar \phi}_-
\left(
(T+\overline{T})^{{n_K-n_i \over 2}+p_{Ki}}Z''_{Ki}W^0_{iLM}
\right)_{\rm symm.}
\right]
\nonumber \ ,
\label{trilinear}
\end{eqnarray}
where {\rm symm.} denotes the symmetrized parts in the (KLM) indices
and
\begin{equation}
W_{KLM}^0 = e^{K \over 2} (K^{-1/2})_K^{K'} (K^{-1/2})_L^{L'}
(K^{-1/2})_M^{M'} W_{K'L'M'} = (T+\overline{T})^{-{(3+n_K+n_L+n_M)
\over 2}} W_{KLM} \label{rescaling}
\end{equation}
are the low-energy (for canonically normalized fields) Yukawa
couplings.

Comments similar to the ones concerning the flavor-dependence of soft
masses apply here. Analogously to the discussion concerning soft
scalar masses, the dependence of trilinear $A$-couplings on the
unknown quantities $n_i, Z'_{i{\bar j}}, Z_{i{\bar j}}^{''} $ can be
neglected under reasonable assumptions.
    For the phenomenological analysis performed in
the next section we analyze in detail the gravity-universal case
$A_{KLM} = 3 m_{3/2} W^0_{KLM}$.
\subsection{$\mu$ and $B_{\mu}$ terms}

The $\mu$ parameter and bilinear coupling arises in our model through
a Giudice-Masiero mechanism \cite{Giudice:1988yz}. We will suppose
a Kahler metric of the form

\begin{equation}
K=K_0 + Z(T,\overline{T})
\left[
H_1 H_2 + \mathrm{h.c}
\right]
\end{equation}

\noindent
where $Z(T,\overline{T})$ is a modular function ensuring
the modular invariance of the term $Z(T,\overline{T})H_1H_2$.
With this convention, we can deduce

\begin{equation}
\mu = m_{3/2} \ Z(T,\overline{T}) \quad , \quad
B_{\mu}=m_{3/2}^2
\left[
2 Z(T,\overline{T}) + G^{\alpha}\nabla_{\alpha} Z(T,\overline{T})
\right]
 + \sum_a g_a^2 D_a \nabla_{H_1} \nabla_{H_2} D_a
\end{equation}

\noindent with $D_a \nabla_{H_1} \nabla_{H_2} D_a = - \frac{3}{2}\xi
D \partial_T Z(T,\overline{T})$ in our case. The modular function
$Z(T,\overline{T}) $ allows a certain flexibility of $\mu$ and $B
\mu$ terms with respect to the gravitino mass. We will use this
flexibility in order to determine the appropriate parameters from
the analysis of electroweak symmetry breaking in the next section.


\subsection{Phenomenology}

If we apply the previous soft term calculations to the numerical
example of Eq. (\ref{num}) we obtain ${\tilde m}_0=3.3$ TeV and
$M_{a}=330$ GeV. In this case, we have a splitting (a factor 10)
between scalar masses ${\tilde m}_0$ and gaugino masses $M_a$, smaller by a factor of two
than in the classical KKLT case. This implies that the one loop
contributions (AMSB) are less important here compared to the
tree-level one. As we already mentioned, this comes from the fact
that $G_T$ participate more actively to the SUSY breaking and
therefore its contribution to the gaugino masses is more important
compared to the KKLT or classical F-term uplifting cases. We show
the spectrum of some typical points in table \ref{tab1}. The
absolute value of $\mu$ is determined by the minimisation condition
of the Higgs potential (assuming CP conservation), but its sign is
not fixed. Furthermore, instead of  $B$ it is more convenient to use
the low energy parameter
 $\tan \beta = \la H_2^0\ra/\la H_1^0\ra$, which is a function of $B$ and the
other parameters.
The low energy mass spectrum  is calculated using the Fortran package {\tt SUSPECT} \cite{Suspect} and  its routines  were described in detail in ref. \cite{Suspect}.
The evaluation of the $b \rightarrow s \gamma$ branching ratio, the anomalous moment of the muon and the  relic neutralino density
is carried out using the routines provided
the program {\tt micrOMEGAs2.0} \cite{micromegas1}.
Minimizing the Higgs potential in the MSSM leads to the standard relation
\begin{equation}
\mu^2 =
\frac{-m_{H_2}^2 {\mathrm {tan^2}}\beta + m_{H_1}^2}
{{\mathrm{tan^2}}\beta -1}
-\frac{1}{2}M_Z^2 \, ,
\label{electroweak}
\end{equation}
This minimization condition is imposed at the scale
$M_{\rm SUSY}=\sqrt{m_{\tilde t_1} m_{\tilde t_2}}$.
Eq. (\ref{electroweak}) can be  approximated in most cases by
\begin{equation}
\label{e.ewa}
\mu^2 \approx - m_{H_2}^2 - \frac{1}{2} M_Z^2
\end{equation}
When the right hand side is negative, electroweak breaking cannot occur.
The Higgs mass parameter  $m_{H_2}^2$ is positive at the GUT scale, but
decreases with decreasing scale down to $M_{SUSY}$, through the contributions
it receives from RG running,
 ${\pa  m_{H_2}^2 \over \pa \log \mu} \approx 6 y_t^2
(m_{H_2}^2 + m_{U_3}^2 + m_{Q_3}^2 + A_t^2)$.
Typically, the value of $m_{H_2}^2$ at the scale $\msusy$ depends mainly on the
soft breaking terms $m_{U3}^2$ and  $m_{Q3}^2$.

\begin{center}
\begin{table}
\centering
\begin{tabular}{|c|c|c|}
\hline
& \bf{A} & \bf{B}   \\
\hline
$\mathrm{W_0 }$ &$-7~10^{-13}$  &  $-4.3~10^{-13}$    \\
$\mathrm{m }$ & $7.3~10^{-12}$ &  $4.5~10^{-12}$   \\
$a$ & 1 & 1   \\
$b$ & 0.3 & 0.5   \\
$q$& 1 & 1   \\
$\tan \beta$ & 30 & 15  \\
$t$ & 98.3 & 59.4  \\
\hline
$\mu$ (GeV) & 810 & 1070   \\
$B\mu$ $(GeV)^2$ & $(403)^2$ & $(871)^2$  \\
\hline
$m_{\chi^0_1}$ & 113 & 145   \\
$m_{\chi^+_1}$ & 224 & 286   \\
$m_{\tilde g}$ & 762 & 948  \\
\hline
$m_{h}$ & 118.7 & 120.1    \\
$m_{A}$ & 2220 & 3291   \\
\hline
$m_{\tilde t_1}$ & 1385 & 1770   \\
$m_{\tilde t_2}$ & 1918 & 2612  \\
$m_{\tilde c_1}, ~ m_{\tilde u_1}$ & 2577 & 3302  \\
\hline
$m_{\tilde b_1}$ & 1913 & 2610  \\
$m_{\tilde b_2}$ & 2313 & 3226  \\
$m_{\tilde s_1}, ~ m_{\tilde d_1}$ & 2578 & 3303   \\
\hline
$m_{\tilde \tau_1}$ & 2288 & 3198   \\
$m_{\tilde \tau_2}$ & 2424 & 3235   \\
$m_{\tilde \mu_1}, ~ m_{\tilde e_1}$ & 2553 & 3275   \\
\hline
\end{tabular}
\caption{Sample spectra. All superpartner masses are in GeV, whereas
$W_0$, $m$ and $t$ are given in Planck units. Both spectra give a
relic density much above the WMAP constraint.} \label{tab1}
\end{table}
\end{center}

\begin{center}
\begin{table}
\centering
\begin{tabular}{|c|c|c|}
\hline
 & \bf{A+GMSB} & \bf{B+GMSB}  \\
\hline
$\mathrm{W_0 }$ &$-7~10^{-13}$  &  $-4.3~10^{-13}$    \\
$\mathrm{m }$ & $7.3~10^{-12}$ &  $4.5~10^{-12}$   \\
$a$ & 1 & 1   \\
$b$ & 0.3 & 0.5   \\
$q$& 1 & 1   \\
$\tan \beta$ & 30 & 15  \\
$t$ & 97.3 & 59.4  \\
\hline
$\lambda$ & $1.7~10^{-3}$ & $1.1~10^{-3}$  \\
$N_{\mathrm{Mess}}$ & 6 & 6  \\
\hline
$\mu$ (GeV) & 186 & 216   \\
$B\mu$ $(GeV)^2$ & $(328)^2$ & $(728)^2$  \\
\hline
$m_{\chi^0_1}$ & 117 & 152   \\
$m_{\chi^+_1}$ & 165 & 201   \\
$m_{\tilde g}$ & 848 & 1057  \\
\hline
$m_{h}$ & 119.7 & 121.1    \\
$m_{A}$ & 1744 & 2769   \\
\hline
$m_{\tilde t_1}$ & 993 & 1225   \\
$m_{\tilde t_2}$ & 1281 & 1713  \\
$m_{\tilde c_1}, ~ m_{\tilde u_1}$ & 1951 & 2417  \\
\hline
$m_{\tilde b_1}$ & 1251 & 1701  \\
$m_{\tilde b_2}$ & 1932 & 2686  \\
$m_{\tilde s_1}, ~ m_{\tilde d_1}$ & 1952 & 2418   \\
\hline
$m_{\tilde \tau_1}$ & 2128 & 2872   \\
$m_{\tilde \tau_2}$ & 2164 & 2963   \\
$m_{\tilde \mu_1}, ~ m_{\tilde e_1}$ & 2292 & 2912   \\
\hline
$\Omega \mathrm{h}^2$ & 0.122 & 0.117   \\
\hline
\end{tabular}
\caption{Sample spectra including gauge mediation contribution.
 All superpartner masses are in GeV, whereas $W_0$,
$m$ and $t$ are in Planck units. The last line correspond to the
relic abundance, within WMAP bounds in each case.} \label{tab2}
\end{table}
\end{center}



\section{Anomalies and Gauge Messengers}

\subsection{Anomalies and messengers}

Anomaly arguments that we discussed in Section 3 and coupling of the
MSSM gauge couplings to the T-modulus introduced in the previous
section strongly suggest that there should be fields carrying Standard
Model quantum numbers charged under the additional $U(1)_X$. Indeed, the
couplings (\ref{sbt1}) generate, through the shift of T under $U(1)_X$
gauge transformation (\ref{model1}) mixed $U(1)_X-G_a^2$ anomalies, with
$G_a = SU(3), SU(2)_L, U(1)_Y$ being a SM group factor. These
anomalies imply some SM-charged fields have to carry {\it positive}
(for positive coefficients $c_a$ in (\ref{sbt1})) $U(1)_X$ charges in
order to cancel, via the 4d Green-Schwarz mechanism, the mixed
anomalies.
There are two generic possibilities that realize this, that we
consider in turn~:

\begin{itemize}

\item The SM quarks and leptons carry $U(1)_X$ charges. In this case,
the squarks and the sleptons will acquire D-term soft masses ${\tilde m}_0^2
\sim D \sim 100$ TeV. If we wish to keep some light superpartners and
to minimize the fine-tuning of the electroweak scale, one possibility
would be to give a charge to the first two generations only
\cite{ckn}. The large hierarchy betwen the first two and the third
generation of squarks can generate various problems, in particular the
third generation could become tachyonic through the RGE running
towards low-energy \cite{ahm}.

\item All MSSM fields (quarks, leptons and Higgses ) carry no $U(1)_X$
  charges. In this case, there should be additional fields carrying
  both SM and $U(1)_X$ charges. In order to preserve perturbative gauge
  coupling unification and be able to give these states a large mass,
we only consider complete vector-like $SU(5)$ multiplets, called
generically $M$ and ${\tilde M}$ in what follows. Notice that these
fields have precisely the features of the so-called "messenger" fields
in gauge-mediation scenarios \cite{Giudice:1998bp}.

\end{itemize}

These arguments strongly suggest therefore to introduce heavy
messengers which can contribute significantly to the soft SUSY
masses breaking terms. In our model, coupling the charged field
$\phi_-$ to the messengers pushes naturally the messenger scale up
to the GUT scale, still giving rise to important contributions to the
scalar masses.

The superpotential is of the form :

\begin{equation}
W_{\mathrm{mess}} = \lambda \phi_- M \tilde M \ , \label{wmess}
\end{equation}

\noindent where $M$ and $\tilde M$ represent the messenger fields of
charges $q$ and $\tilde q$ respectively. Without loss of generality,
we will take $q=\tilde q= + 1/2$ thorough the rest of the analysis.
Notice that the messenger fields, vector-like wrt to SM gauge
interactions, are chiral wrt the anomalous $U(1)_X$ symmetry. In
(\ref{wmess}), $\lambda$ is the low-energy coupling, related to the
high-energy supergravity coupling by a formula similar to
(\ref{rescaling}). Notice that for zero (or positive) modular
weights for $\Phi_-$, $M$ , ${\tilde M}$, the low energy coupling
$\lambda$ is highly suppressed wrt the high-energy ones by inverse
powers of $T+{\bar T}$.

In general, adding messengers to a supersymmetry breaking sector
generates a new, supersymmetry preserving vacuum. This is because in
order to generate gaugino masses we have to explicitly break
R-symmetry, which in turns generically restores supersymmetry \cite{nelson}.
In our case, however, due to the presence of the $U(1)_X$ gauge
symmetry, this does not happen~; even in the presence of messenger
fields, there is no supersymmetry preserving vacuum.
 This is an important difference compared to standard gauge mediation models of
supersymmetry breaking.

Another very important outcome of the charged nature of messenger
fields is a new D-term contribution to
scalar messenger masses.

\noindent
The scalar messenger mass matrix is
\begin{equation}
M_{\mathrm{mess}}^2=
\left(
\begin{array}{cc}
(\lambda \phi_-)^2 + \frac{1}{2} g_X^2 D & \lambda F_- \\
\lambda F_- & (\lambda \phi_-)^2 + \frac{1}{2} g_X^2 D
\end{array} \right)
\end{equation}

\noindent
Once diagonalized the messenger scalar mass matrix, the two eigenvalues
are :

\begin{equation}
m_-^2= \left[(\lambda \phi_-)^2 + \frac{1}{2} g_X^2 D\right] - \lambda F_- ~~~~~~~
m_+^2= \left[(\lambda \phi_-)^2 + \frac{1}{2} g_X^2 D\right] + \lambda
F_- \ ,
\end{equation}

\noindent whereas the fermion mass is given by :

\begin{equation}
m_f \ = \ \lambda \phi_- \ .
\end{equation}

Notice that
\begin{equation}
(Str M^2)_{\rm mess.} \ = \ 2 \ g_X^2 \ D \ \not = \ 0 \ . \label{mess1}
\end{equation}

By standard gauge-mediation type diagrams, gaugino masses are induced
at one-loop, whereas scalar masses are induced at two-loops. Due to (\ref{mess1}),
the computation of the scalar masses is slightly different compared to
the standard gauge-mediation models, as shown by Poppitz and Trivedi
\cite{Poppitz:1996xw}. In particular the result is not anymore UV
finite, there is a logarithmically divergence term which will play a
crucial role in what follows.

In the context of our model, the uplift relation (\ref{upliftW0}) has
very strong phenomenological implications.
Indeed, if D-term contributions appear in the scalar soft mass
terms of the visible sector, through the two loops--suppressed GMSB
mechanism , it turn out that their magnitude is automatically of the
same order as the gravity (i.e. $m_{3/2}$) contribution.
This is clearly seen from our numerical example (\ref{num}), in
particular from the values of the $D$-term.\\
\subsection{Soft masses}

The exact calculation of the radiatively induced gaugino and
scalar masses is performed
in \cite{Poppitz:1996xw,Giudice:1998bp}.
For one messenger multiplet, we obtain for the gaugino mass

\begin{equation}
M^{\mathrm{GMSB}}_a=\frac{g_a^2 m_f S_Q}{8 \pi^2} \
\frac{y_- \log{y_-}-y_+\log{y_+}-y_- y_+ \log{(y_-/y_+)}}{(y_--1)(y_+-1)}
\end{equation}

\noindent
and for the scalar masses

\begin{equation}
({\tilde m}_0^{\mathrm{GMSB}})^2 \ = \ \sum_a \frac{g_a^4}{128 \pi^4}m_f^2 C_a S_Q
 \ F(y_-,y_+,\Lambda_{\mathrm{UV}}^2/m_f^2)
\end{equation}

\noindent
with $y_i=m_i^2/m_f^2$ and
where $g_a$ is the corresponding SM gauge coupling (unified at high scale),
$C_a$ is the Casimir in the MSSM scalar fields representations
(normalized as $C_a (N) = (N^2-1)/(2N)$ for the fundamental
representation of $SU(N)$ gauge group,  while for $U(1)_Y$
it is simply $Y^2$ )
and $S_Q$ the Dynkin index of the messenger representation
(normalized to 1/2 for a fundamental of $SU(N)$). The function $F$ is given by

\begin{eqnarray}
F(y_-,y_+,\Lambda_{\mathrm{UV}}^2/m_f^2)&=&
-(2 y_-+2y_+-4)\log{\frac{\Lambda^2_{\mathrm{UV}}}{m_f^2}}
\nonumber
\\
&+& 2(2y_-+2y_+-4) +(y_-+y_+)\log{y_-}\log{y_+}
\nonumber
\\
&+& G(y_-,y_+) + G(y_+,y_-) \ ,
\end{eqnarray}

\noindent
where

\begin{eqnarray}
G(y_-,y_+)&=&  2y_-\log{y_-}+(1+y_-) \log^2{y_-}-\frac{1}{2}(y_-+y_+)\log^2{y_-}
\nonumber
\\
&+& 2(1-y_-)\mathrm{Li_2}(1-\frac{1}{y_-}) + 2 (1+y_-)\mathrm{Li_2}(1-{y_-})
\nonumber
\\
&-& y_- \mathrm{Li_2}(1-\frac{y_-}{y_+}) \ .
\end{eqnarray}

\noindent
$\mathrm{Li_2}(x)$ above refers to the dilogarithm function and is defined
by $\mathrm{Li_2}(x)= - \int_0^1 dz z^{-1}\log{(1-xz)}$.
After an expansion in the perturbative parameter
$\epsilon=\lambda F_-/(\lambda \phi_-)^2$ the mass terms become

\begin{equation}
M^{\mathrm{GMSB}}_a=S_Q \frac{m_0 g_a^2}{8 \pi^2}
\left(
\frac{\phi_+}{\phi_-}
\right) \ ,
\label{gauginoGMSB}
\end{equation}

\noindent
where $m_0$ is the low-energy mass parameter of the FI model, equal to
$m_0 = m / (T+{\bar T})^{3/2}$ for our model in Section 3, and

\begin{eqnarray}
({\tilde m}^{\mathrm{GMSB}}_0)^2 &=& \sum_a
\frac{g_a^4}{128 \pi^4} C_a S_Q
\left[
-2 g_X^2D \log\left(
\frac{\Lambda_{\mathrm{UV}}}{\lambda \phi_-}
\right)^2 + 2 g_X^2D +G(y_-,y_+)+G(y_+,y_-)
\right]
\nonumber
\\
&=&
\frac{m_0^2}{64 \pi^4}
\sum_a g_a^4 C_a S_Q
\left[
1- \log
\left(
\frac{\Lambda_{\mathrm{UV}}}{\lambda \phi_-}
\right)^2
+\left(
\frac{\phi_+}{\phi_-}
\right)^2
\right] \ , \label{scalarGMSB}
\end{eqnarray}
where in the last line we used (\ref{D}).

One important feature of Eq.(\ref{scalarGMSB}) is the presence of
the $\log(\Lambda_{\mathrm{UV}} / \lambda \phi_-)$ term in the soft
scalar masses. This logarithmic divergence arises typically in the
presence of anomalous $U(1)_X$ that gives a non-vanishing supertrace
(\ref{mess1}) for the messengers superfields \cite{Poppitz:1996xw}.
In low energy-GMSB, it usually limits the scale beyond which "new
physics" occurs, because the scalars become tachyonic already for
$\Lambda_{\mathrm{UV}} / \lambda \phi_-$ around 50. In our specific
case, the running is much shorter : from the FI scale ($\phi_-$)
to the Planck scale (a factor less than 10).

Some remarks are in order concerning the anomaly-mediation
contribution to soft terms. For scalar masses, they are completely
negligible compared to both gravity and the non-standard GMSB
contributions (\ref{scalarGMSB}). For gaugino masses, they are much smaller than the
gravity contribution, whereas they are suppressed wrt to the standard
GMSB contributions (\ref{gauginoGMSB}) only by the number of messenger
fields $1/N_{\rm mess}$. Since we consider relatively large values $N_{\rm mess}=6$
in our analysis, we can neglect also the anomaly contributions to
gaugino masses in what follows.
\subsection{Phenomenological effects}

In the complete model, scalar and gaugino masses get contributions both from
gravity and the gauge mediation diagrams
\begin{eqnarray}
&& ({\tilde m}_0^2)_{} \ = \ ({\tilde m}_0^2)_{\rm grav.} \ + \ N_{\rm Mess}
({\tilde m}^{\mathrm{GMSB}}_0)^2 \ , \nonumber \\
&&  M_a \ = \ (M_a)_{\rm grav.} \ + \ N_{\rm Mess}
(M^{\mathrm{GMSB}}_a) \ . \label{pheno1}
\end{eqnarray}

The negative contribution to the scalar masses ${\tilde m}_0$ induced by the ultraviolet
divergence has strong consequences on the mass spectrum and the
phenomenology of the model. It reduces significantly the masses in the
left--handed squark sector (the more charged under the SM gauge group)
and can have repercussion in the neutralino sector through
$M_{H_1}$. In addition, decreasing the value
of $m_{U3}^2$ and  $m_{Q3}^2$ with gauge mediation naturally decreases
the value of $\mu^2$ through Eq.(\ref{e.ewa}).

We show in Tab.(\ref{tab2}) (Tab.(\ref{tab1}))  the spectrum with
(without) the gauge mediation contributions, after including the RG
evolution to low-energy. The scalar spectrum and nature of
neutralino (through the $\mu$ parameter) are considerably altered.
On the other hand, the positive GMSB contribution to gauginos
compresses even more the supersymmetric spectrum, especially for a
large number $N_{\rm Mess}$ of messenger fields. Notice that,
whereas for traditional messenger masses (i.e. around $100-1000$
TeV), the RG running up to the unification scale forces $N_{\rm
Mess} \le 3$ in order to avoid strong coupling effects, in our case
since messengers have masses of order $10^{17}$ GeV, the number of
messengers can be larger. By using this (and/or also the alternative
possibility of enhancing the negative contribution to scalar masses
by decreasing the coupling $\lambda$) we can obtain the efficiently
compressed spectrum displayed in Table 2.

\begin{figure}
    \begin{center}
\centerline{
       \epsfig{file=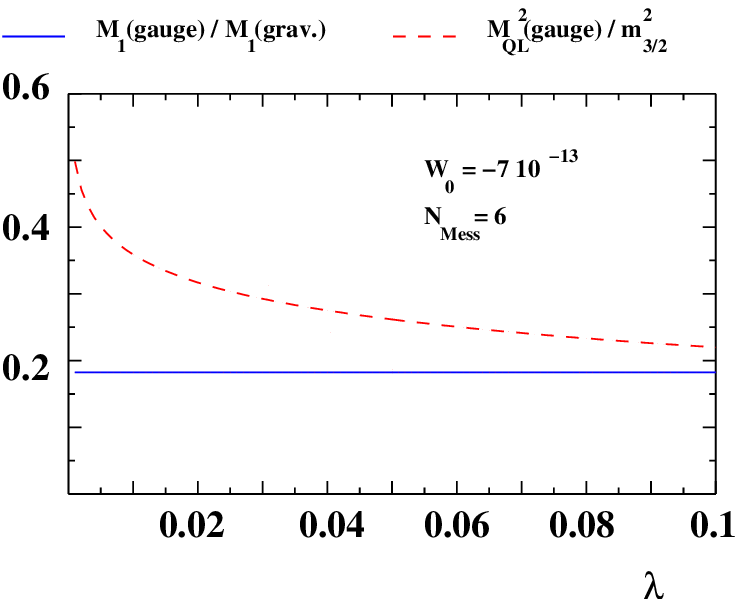,width=0.5\textwidth}
\hskip 1cm
       \epsfig{file=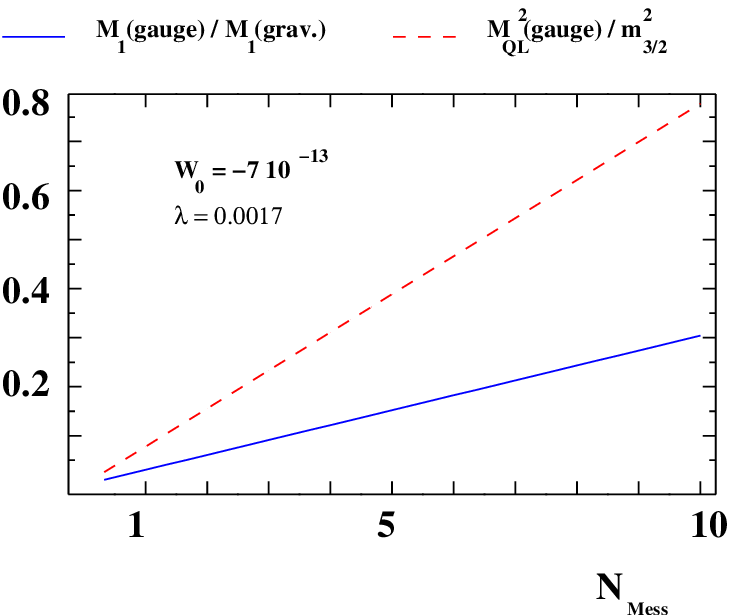,width=0.5\textwidth}
       }
          \caption{{\footnotesize
Ratio of the (gauge mediated)/(gravity mediated) for the gaugino
masses (blue line) and the scalar masses (red dashed line) as
function of $\lambda$ (left) and the number of messenger (right). }}
\label{fig:ratio}
\end{center}
\end{figure}

Notice that, in contrast to other scenarios (see for example
\cite{Nomura:2007cc} ) where the gauge/gravity relative
contributions are completely fixed, in our case, due to the presence
of the two charged fields $\Phi_{\pm}$, the gauge and the gravity
contributions to soft terms are governed by different parameters. It
is instructive to see in Figs.(\ref{fig:ratio}) the dependence  of
the gauge contribution to soft terms as a function of the relevant
parameters of the model ($\lambda$ and $N_{\rm Mess}$). For low
values of $\lambda$, the gauge contribution to the scalar mass
becomes important, and even of the same order of magnitude than the
gravity contribution for $\lambda \sim 10^{-3}$. Indeed, smaller
values of $\lambda$ implies lighter messenger and thus a larger
running between $M_{\mathrm{mess}}$ and
$\Lambda_{\mathrm{\mathrm{UV}}}$. Gaugino masses are not affected by
$\lambda$. The number of messenger acts directly on the scalar and
gaugino masses, and the gauge contribution becomes relevant in both
cases for $N_{\rm Mess} \sim 6$.

\section{Conclusions}

In this work, we tried to combine the various ingredients that a
microscopic string theory can provide in order to successfully
stabilize moduli fields with a  TeV gravitino mass. This is realized
via a supergravity version of the Fayet-Iliopoulos (FI) model with
an anomalous $U(1)_X$ gauge symmetry, in which a (non-linearly
charged) modulus field $T$ plays an instrumental role, whereas in
turn the FI sector plays a crucial role in supersymmetry breaking
and for getting a zero vacuum energy. Due to the non-decoupling
between the modulus and the uplift sector, the contribution of the
modulus $T$ to supersymmetry breaking is higher than in previous
schemes. This increases the numerical values of the gaugino mass and
renders less important the loop contributions to soft terms.

Due to the intrincate definition of the model, anomaly arguments
strongly suggests the presence in the spectrum of charged fields
which have properties similar to messenger fields of gauge mediation
of supersymmetry breaking. Due to the charged nature of our
messengers, their superpotential couplings are R-symmetric. As a
consequence, in contrast to standard gauge mediation scenarios, our
messengers do not restore supersymmetry~; there is no new
supersymmetric vacuum state due to their presence and couplings. Our
model has therefore a completely stable non-supersymmetric ground
state, which is difficult to realize in more standard
gauge-mediation scenarios.  Whereas, for the standard reasons, our
messengers are vector-like with respect to SM gauge interactions,
due to anomaly cancellations they are however chiral with respect to
the $U(1)_X$ interactions. Consequently, due to their coupling to
the FI supersymmetry breaking sector, they have a particular
spectrum, in particular $(Str M^2)_{\rm mess.} \sim D \not=0$. The
resulting mixed gravity/gauge mediation scenario is therefore of
non-standard type~: the two-loop gauge mediation contributions to
MSSM scalars are negative \cite{Poppitz:1996xw,amm}.

 The mass scales in the problem are such that gravity and gauge
mediation contributions to scalar soft masses are comparable and
compete with each other, providing an original predictive spectrum
and phenomenology. Indeed, squarks, which are the heaviest
superpartners in most mediation scenarios, are here typically the
lightest scalars since they get the biggest negative contributions
from the non-standard gauge-mediation contribution
(\ref{scalarGMSB}). Taking into account the larger than usual
gravity-induced gaugino masses and the additional positive
contribution to them coming from gauge mediation diagrams, we end up
with an original low-energy spectrum in which the whole
superpartners spectrum is more compressed than in the usual mSUGRA,
gauge or mixed modulus-anomaly mediation scenarios. The complete
mixed model of Section 6 has the Higgsinos as the LSP and a good
relic abundance, compatible with the WMAP bounds. The peculiar
details of the spectrum, like the universality of gaugino masses at
the unification scale and the negative GMSB contribution to scalar
masses, rendering squarks lighter than sleptons, could be tested at
LHC and certainly deserve a more focused study.

Finally, whereas in the present paper we get a bigger modulus
contribution to supersymmetry breaking than in previous
O'Rafeartaigh type models, the main contribution still comes from
the uplift sector and more precisely in our case from the $\Phi_+$
field. It is a very interesting and open question to find explicit
realizations, with complete moduli stabilization and zero vacuum
energy, of the string-inspired supersymmetry breaking
parametrizations \cite{bim}, which assumes moduli/dilaton
domination. To our knowledge, this does not seem to be realized in
the current known models of moduli stabilization.

\section{Appendix~: Dynamical origin for the superpotential}

The aim of this appendix is to justify the coupling
\begin{equation}
W = ... + a e^{-bT} \phi_{-}^{q} +...
\end{equation}
that we considered in the superpotential of our model. We will show that this
term has its origin in strong coupling regime effects for the non-abelian
gauge group of the hidden sector, and that in particular it is induced by
"integrating out" the mesonic fields which are the right degrees of
freedom describing the theory in such regime.\\
As discussed in section 4, the microscopic description of the
model implies $N_f < N$ chiral multiplets in the   representions $Q=(N,1)$
and $\tilde{Q} = (\bar{N},1)$ of the gauge group $U(N) \times U(1)_X$.
Naturally, at a characteristic scale
\begin{equation}
\Lambda = M_P ~ e^{- \frac{2 \pi}{3N - N_f}~T} \ , 
\end{equation}
the non-abelian gauge group will enter in a strong coupling regime, it
will undergo a gaugino condensation and the model is properly described in
terms of the mesonic field $M=Q\tilde{Q}$, of generic charge $q'$ under
the $U(1)_X$ gauge group (once normalized at $-1$ the charge of the field
$\phi_{-}$). A non-perturbative ADS potential is generated:
\begin{equation}
W_{np} = ( N - N_f) ~\left( \frac{\Lambda^{3N - N_f}}{det M}
\right)^{\frac{1}{N - N_f}} \ . 
\end{equation}
Including the most general disk-level perturbative open string coupling,
and the coupling between $\phi_{-}$ and $\phi_{+}$ discussed in section
4, the gauge invariant superpotential reads \footnote{For the
aim of this appendix, the term $W_0$ is completely irrelevant.}
\begin{equation}
W =   ( N - N_f) ~\left( \frac{\Lambda^{3N - N_f}}{det M}
\right)^{\frac{1}{N - N_f}} + \left( \frac{\phi_{-}}{M_P} \right)^{q'}
\lambda_{i}^{\bar{j}}M_P~ M_{\bar{j}}^{i} +m \phi_{+} \phi_{-} \ . 
\end{equation}
The auxiliary fields and the D-term can now be calculated
\cite{Binetruy:1996uv}:
\begin{eqnarray}
\left( \bar{F}_{M^{\dagger}} \right) ^{\bar{i}}_{i}&=&2\left[  - \left(
M^{-1}\right)^{\bar{j}}_{i} \left(\frac{\Lambda^{3N - N_f}}{det M}
\right)^{\frac{1}{N - N_f}} + \left( \frac{\phi_{-}}{M_P} \right)^{q'}
\lambda_{i}^{\bar{j}} M_P  \right]~\left[ \left(
M^{\dagger}M\right)^{\frac{1}{2}}\right]^{i}_{\bar{j}} \ , \nonumber \\
\bar{F}_{\bar{\phi}_{-}}&=& q'\left( \frac{\phi_{-}}{M_P} \right)^{q'-1}
Tr\left( \lambda M \right) + m \phi_{+} \ , \nonumber \\
\bar{F}_{\bar{\phi}_{+}}&=& m \phi_{-} \ , \nonumber \\
D &=& \left[ q' Tr\left( M^{\dagger} M\right)^{1/2} - |\phi_{-}|^2 +
|\phi_{+}|^2 + \frac{\xi^2}{T+\bar{T}} \right] \ .
\end{eqnarray}

From a simple analysis of the equations of motions for the mesons,
$\phi_{-}$ and $\phi_{+}$, it is possible to see that in the
minimum, under the conditions
\begin{equation} \label{cond1}
\Lambda^2 < m^2 \ll \frac{\xi^2}{T+\bar{T}} < M_P^2
\end{equation}
and in particular requiring \footnote{This in order to assure that
$|\langle \phi_{+} \rangle| \ll |\langle \phi_{-} \rangle|$,in accord with
what happens in the usual FI global model.}
\begin{equation} \label{cond2}
q' N_f \left( det \lambda \right)^{\frac{1}{N}} \left(
\frac{M_P}{m}\right) \left( \frac{\langle \phi_{-} \rangle}{M_P}
\right)^{q' \frac{N_f}{N}-2} \left( \frac{\Lambda}{M_P} \right)^{\frac{3N
- N_f}{N}} \ll 1 \ , 
\end{equation}
the contribution of the D-term is negligible, $\langle \phi_{-}
\rangle^2 \sim \frac{\xi^2}{T+\bar{T}}$ and the value of the F-terms
for the mesons are very small. These F-terms satisfy the relation
\begin{equation}
\frac{\langle F_{M} \rangle}{ \langle M \rangle } \sim \frac{ \langle
F_{\phi_{-} \rangle} }{ \langle \phi_{-} \rangle} \sim q' N_f \left( det
\lambda \right)^{\frac{1}{N}}~ \left( \frac{\Lambda}{M_P}\right)^{\frac{3N
- N_f}{N}}\left(\frac{\langle \phi \rangle}{M_P} \right)^{\frac{q'
N_f}{N}-2} ~M_P \ . 
\end{equation}
Since in the potential the contributions of these F-terms are
respectively proportional to $|\langle F_{M} \rangle|^2 /| \langle M
\rangle|$ and $|\langle F_{\phi_{-}} \rangle|^2$, as long as the
vev's for the mesons $| \langle M \rangle|$  is very small compared
to $|\langle \phi_{-} \rangle|^2$,
 we are therefore allowed to integrate out the mesons $M^{i}_{\bar{j}}$
and the effective superpotential we obtain has the form
\begin{equation} \label{effpot}
W^{eff} = N  ~\left( \frac{\Lambda}{M_P}\right)^{\frac{3N -
N_f}{N}}~\left( \frac{\phi_{-}}{M_P} \right)^{\frac{q' N_f}{N}}
~\left( det \lambda \right)^{\frac{1}{N}} ~M_P^3 +m \phi_{+}
\phi_{-} \ .
\end{equation}
Nonetheless, as one can show resolving in the first approximation the
equation $\bar{F}_{M^{\dagger}} = 0$, the vev of the meson
$M^{i}_{\bar{j}}$ is approximatively
\begin{equation}
\langle M^{i}_{\bar{j}} \rangle \sim N_f (\lambda^{-1})^{i}_{\bar{j}}
(det\lambda)^{\frac{1}{N}} \left( \frac{\langle \phi_{-} \rangle}{M_P}
\right)^{q' \frac{N_f -N}{N}} \left( \frac{\Lambda}{M_P} \right)^{\frac{3N
- N_f}{N}} M_P^2
\end{equation}
and then, using (\ref{cond2}) and $\lambda_{i}^{\bar{j}} \sim
\delta_{i}^{\bar{j}}$, we have
\begin{equation} \label{limitsonq}
\frac{|\langle M \rangle|}{|\langle \phi_{-} \rangle|^2} \ll \left(
\frac{m}{M_P} \right) \left( \frac{\langle \phi_{-} \rangle}{M_P}
\right)^{-q'} \ .
\end{equation}
Therefore, for $m$ small enough (\ref{cond1}) and for reasonable
$q'$, this ratio is actually $\ll1$ and the "integration out" is
consistent. By re-writing the superpotential  (\ref{effpot}) by
using the definitions
\begin{eqnarray}
a &=& N  \left(det \lambda \right)^{\frac{1}{N}} \ , \nonumber \\
e^{-b T} &=& \left( \frac{\Lambda}{M_P}\right)^{\frac{3N - N_f}{N}} \
, \nonumber \\
q &=& \frac{q' N_f}{N} \ ,
\end{eqnarray}
we find exactly the form of the superpotential used in the equation
(\ref{Eq.W}) once the right powers of $M_P$ restored.\\
As a numerical example, in order to check if the results obtained in the
paper agree with a reasonable nonperturbative scale $\Lambda$, we can
consider the special case $N= 2, ~N_f = 1,~ q'=2,~ \lambda_{i}^{\bar{j}}
\sim \delta_{i}^{\bar{j}}~$, i.e. the case studied numerically in the
section 5.5. With the choice of the parameters done in that
example, we can evaluate
\begin{equation}
\Lambda^{5/2} \sim a e^{-bt} M_P^{5/2} \sim 10^{-13} M_P^{5/2} \ , 
\end{equation}
which means that in this scenario we expect that the non-abelian
gauge group $U(N)$ enters in a strong coupling regime at a sensible
scale of order $\Lambda \sim 10^{14}$ GeV. Moreover, we can check if
in this case the approximations done for the "integration out" step
are good. Actually, since $m < 10^{-11} M_P $ and $\langle \phi_{-}
\rangle \sim 10^{-1} M_P$, it is clear that (\ref{limitsonq}) is
verified and that therefore the whole procedure is consistent.

\section*{Acknowledgments}{ Work partially supported by the CNRS PICS \#~2530
and 3059, RTN contracts MRTN-CT-2004-005104 and MRTN-CT-2004-503369,
the European Union Excellence Grant, MEXT-CT-2003-509661 and the
European contract MTKD-CT-2005-029466. E.D. would like to thanks the
Institute for Theoretical Physics of Warsaw form warm hospitality and
financial support via the "Marie Curie Host Fellowship for Transfer of
Knowledge" MTKD-CT-2005-029466. The work
of Y.M. is sponsored by the PAI programm PICASSO under contract
PAI--10825VF. He would like to thank the European Network of
Theoretical Astroparticle Physics ILIAS/N6 under contract number
RII3-CT-2004-506222 and the French ANR project PHYS@COLCOS for
financial support. The work of A.R. was partially supported by INFN
and by the European Commission Marie Curie Intra-European
Fellowships under the contract N 041443. The authors want to thank
Chlo\'e Papineau, Erich Poppitz, Mariano Quiros and Michele Trapletti
for useful discussions. E.D. and S.P. thank the Theory Group of CERN
for hospitality during the completion of this work.}


\nocite{}
\bibliography{bmn}
\bibliographystyle{unsrt}

\end{document}